\documentclass{iopconfser}
\usepackage{amsmath}
\usepackage{graphicx}
\usepackage{hyperref}
\newcommand\mcmule{{{\sc McMule}}}
\newcommand{\eik}{\mathcal{E}}
\newcommand{\ieik}{\hat{\mathcal{E}}}
\newcommand{\M}[2]{\mathcal{M}_{#1}^{(#2)}}
\newcommand{\fM}[2]{\mathcal{M}_{#1}^{(#2)f}}
\newcommand{\D}{\mathrm{d}}
\newcommand{\cell}{\mathcal{C}}

\begin{document}

\title{{\sc McMule} -- a Monte Carlo generator for low energy processes}

\renewcommand*{\thefootnote}{\fnsymbol{footnote}}
\author{Yannick Ulrich$^{1}\footnotemark{}$}
\footnotetext{Affiliation at time of publishing: Department of Mathematical Sciences, University of Liverpool}

\affil{$^1$Institut für Theoretische Physik \& AEC, Universit\"at Bern, Sidlerstrasse 5, CH-3012 Bern, Switzerland}

\email{yannick.ulrich@liverpool.ac.uk}

\begin{abstract}
    \mcmule{}, a Monte Carlo for MUons and other LEptons, implements many major QED processes at NNLO (eg. $ee\to ee$, $e\mu\to e\mu$, $ee\to\mu\mu$, $\ell p\to\ell p$, $\mu\to\nu\bar\nu e$) including effects from the lepton masses, making it suitable for predictions for low-energy experiments such as MUonE, CMD-III, PRad, or MUSE.

    Recently, \mcmule{} gained the ability to generate events at NNLO rather than just pre-defined differential distributions.
    To avoid negative event weights, it employs cellular resampling directly as part of the generation step which further reduces the fraction of negative weights.
\end{abstract}

\section{Introduction}

Higher-order QED corrections have become important for a number of measurements.
Examples include luminosity measurements at $e^+e^-$ colliders, tau physics, and measurements of the proton radius.
Another central use case is the extraction of the hadronic vacuum polarisation contributions to the anomalous magnetic moment of the muon either in $e^+e^-\to\text{hadrons}$ or in $e\mu\to e\mu$.
In the latter case, the MUonE experiment has an unprecedented precision goal of $10^{-5}$.
Finally, searches for New Physics in lepton decays, most notably $\mu\to e\gamma$ and $\mu\to eee$, require very good modelling of the Standard Model (SM) background.

In these proceedings, we will review \mcmule{}~\cite{Banerjee:2020rww}, a Monte Carlo for MUons and other LEptons.
We have implemented all leptonic $2\to2$ processes in QED at next-to-next-to-leading order (NNLO) as well as a few others; a full list is shown in Table~\ref{tab:processes}.
Once provided with the required matrix elements, the \mcmule{} framework can calculate a physical cross section for an arbitrary physical, i.e. IR finite, observable.
In doing so, it takes care of the phase space generation, subtraction~\cite{Engel:2019nfw}, stabilisation~\cite{Banerjee:2021mty}, and integration.

The current stable version of \mcmule{} {\tt v0.5.0} is an integrator which means it can calculate arbitrary distributions but not generate events.
An event generator is currently still in testing as we will discuss below.
The code can be obtained from\\[1em]
\begin{indent}
    \setlength\parindent{24pt}
    \url{https://mule-tools.gitlab.io}
\end{indent}\\[1em]
and the manual, which includes a getting started section, can be found at\\[1em]
\begin{indent}
    \setlength\parindent{24pt}
    \url{https://mcmule.readthedocs.io}
\end{indent}\\[1em]
The code is written in Fortran 95, compiled with {\tt meson+ninja}, and supported by tooling in {\tt python 3.9}.
The user specifies their observable in a routine that is compiled separately and loaded at runtime, allowing them to use whatever language and framework they prefer.

These proceedings are organised as follows:
in Section~\ref{sec:theory}, we will very briefly review some of the theory behind \mcmule{} to the extent that is relevant for its implementation.
Next, we will discuss how \mcmule{}'s event generation mode will work in Section~\ref{sec:generator} with some results for $\mu\to\nu\bar\nu e$ at NNLO before concluding in Section~\ref{sec:conclusion}.

\begin{table}
    \centering
    \begin{tabular}{l|l|l|l|l}
        \bf process                & \bf experiment                & \bf physics motivation            & \bf order  & \bf reference \\\hline
        $e\mu\to e\mu$             & MUonE                         & HVP to $(g-2)_\mu$                & NNLO$+$    & \cite{Banerjee:2020rww,Broggio:2022htr}\\
        $\ell p\to \ell p$         & P2, Muse, Prad, QWeak, ...    & proton radius, weak charge     & NNLO       & \cite{Banerjee:2020rww,Engel:2023arz}\\
        $e^-e^-\to e^-e^-$         & Prad 2                        & normalisation                     & NNLO       & \cite{Banerjee:2021qvi}\\
                                   & MOLLER, ...                   & $\sin^2\theta_W$ at low $Q^2$     &            & \\
        $e^+e^-\to e^+e^-$         & any $e^+e^-$ collider         & luminosity measurement            & NNLO       & \cite{Banerjee:2021mty}\\
        $ee\to\ell\ell$            & VEPP, BES, Daphne, ...        & $R$-ratio                         & NNLO$\pm$  & \cite{Kollatzsch:2022bqa}\\
                                   & Belle                         & $\tau$ properties                 &            & \\
        $ee\to\gamma\gamma$        & Daphne                        & dark searches                     & NNLO$-$    & \\
                                   & any $e^+e^-$ collider         & luminosity measurement            &            & \\
        $e\nu\to e\nu$             & DUNE                          & flux \& $\sin^2\theta_W$          & NNLO$-$    & \\
        $\mu\to\nu\bar\nu e$       & MEG                           & ALP searches                      & NNLO$+$    & \cite{Banerjee:2022nbr}\\
                                   & DUNE                          & beam-line profiling               &            & \\\hline
        $\mu\to\nu\bar\nu e\gamma$ & MEG, Mu3e, Pioneer            & background                        & NLO        & \cite{Pruna:2017upz}\\
        $\mu\to\nu\bar\nu eee    $ &      Mu3e                     & background                        & NLO        & \cite{Pruna:2016spf}\\\hline
        $ee\to\pi\pi$              &      VEPP, BES, Daphne, ...   &       $R$-ratio                   & NNLO$-$    & \\
        $ee\to\ell\ell\gamma$      &      VEPP, BES, Daphne, ...   &       $R$-ratio                   & NLO$+$
    \end{tabular}

    \caption{
        Processes included in \mcmule{}, the experiment and physics case for which they are relevant, and the order at which they are implemented.
        A $+$ indicates that future improvements are to be expected for this process, while a $-$ indicates that this process is not known exactly at the given order.
    }
    \label{tab:processes}

\end{table}

\section{Theory behind \mcmule{}}\label{sec:theory}

To calculate a process in \mcmule{}, it needs to know the relevant matrix elements (squared).
Using $\M n\ell$ to denote the $\ell$-loop $n$-particle matrix element, we need at
\begin{center}
\begin{tabular}{ll}
    LO & $\M n0$,\\
    NLO & $\M n1$ and $\M{n+1}0$,\\
    NNLO & $\M n2$, $\M{n+1}1$, and $\M{n+2}0$.
\end{tabular}
\end{center}
These are represented as function pointers in \mcmule{} and can come from any source.
For the real-virtual $\M{n+1}1$, we usually rely on OpenLoops~\cite{Buccioni:2017yxi,Buccioni:2019sur} supplemented with an automatic stabilisation technique called next-to-soft (NTS) stabilisation~\cite{Banerjee:2021mty} in numerically delicate situations.
\mcmule{} is further able to automatically construct the necessary counter terms and perform the numerical integration using the FKS$^\ell$ subtraction scheme.

\subsection{The \texorpdfstring{FKS$^\ell$}{FKSl} subtraction scheme}

The calculation of higher-order corrections includes divergent real corrections.
Since \mcmule{} utilises numerical integration, we need to include a prescription to handle these in dimensional regularisation.
Since we always consider fermions massive, the only source of singularities are soft emissions, where we can use the universal behaviour studied by Yennie, Frautschi, and Suura (YFS)~\cite{Yennie:1961ad}:
a real-emission matrix element with any number of loops can be approximated in the limit of the emitted photon becoming soft as
\begin{align}
    \M{n+1}\ell = \eik \M{n}\ell + \mathcal{O}(\xi^{-1})\,,
    \label{eq:soft}
\end{align}
where we have defined the rescaled photon energy $\xi = 2E_\gamma/\sqrt{s}$.
We have further defined the eikonal factor $\eik$
\begin{align}
    \eik = -\sum_{ij}Q_iQ_j \frac{p_i\cdot p_j}{(p_i\cdot p_\gamma)(p_j\cdot p_\gamma)}\,,
\end{align}
that considers all pairs of fermions with momenta $p_i$ (assumed incoming) and charges $Q_i$.
Integrating $\eik$ over the phase space of the soft, unresolved photon results in the integrated eikonal $\ieik$, which contains an explicit $1/\epsilon$ soft singularity.
The YFS theorem further states that soft singularities exponentiate such that all singularities from loop integration can be subtracted by $\ieik$ as follows
\begin{align}
    e^{\ieik} \sum_{\ell=0}^\infty \M n\ell = \sum_{\ell=0}^\infty \fM n\ell = \text{finite}\,.
    \label{eq:softvirt}
\end{align}
We can use~\eqref{eq:soft} and~\eqref{eq:softvirt} to construct an all-order extension of the FKS subtraction scheme~\cite{Frixione:1995ms,Frederix:2009yq} called FKS$^\ell$~\cite{Engel:2019nfw} that only requires knowledge of $\eik$ and $\ieik$ which allows for very efficient implementation.
In particular, we have
\begin{align}
    \sigma^{(\ell)} &= \sum_{j=0}^\ell \int\D\Phi_{n+j}\frac1{j!}
        \Bigg[\prod_{i=1}^j \Big(\frac1{\xi_i}\Big)_c\Bigg]
        \fM{n+j}{\ell}\\
    \int_0^1\D\xi \Big(\frac1{\xi_i}\Big)_c &\equiv \int_0^1\D\xi\frac{f(\xi)-f(0)\theta(\xi_c-\xi)}{\xi}\,,
\end{align}
where we have also defined an unphysical parameter $0<\xi_c\le1$ that can be chosen arbitrarily.

\subsection{NTS stabilisation}
To improve the numerical stability and performance when a real photon becomes soft, we utilise NTS stabilisation.
The basic idea is to expand the matrix element $\M{n+1}1$ in the photon energy $\xi$ up to next-to-leading power (NLP) and use the expanded result rather than the full OpenLoops calculation.
While the leading power (LP) term is just the eikonal of~\eqref{eq:soft}, the next term is more complicated.
It was first studied by Low~\cite{Low:1958sn}, Burnett, and Kroll~\cite{Burnett:1967km} at tree level as the LBK theorem and later extended first to one-loop~\cite{Engel:2021ccn} and later to any number of loops~\cite{Engel:2023ifn} and photons~\cite{Engel:2023rxp}.
The case relevant for the real virtual is
\begin{align}
    \M{n+1}1 = \underbrace{\eik\times\M n1}_{\rm LP} + \underbrace{
        D_{\rm LBK}\Big[\M n1\Big]
        + \mathcal{S} \times \M n0
    }_{\rm NLP} + \mathcal{O}(\xi^0)\,.
    \label{eq:nts}
\end{align}
For the exact definitions of the LBK operator $D_{\rm LBK}$ and the soft function $\mathcal{S}$, see for example~\cite{Engel:2023ifn}.
The former is a differential operator that is acting on the one-loop matrix element $\M n1$.
This operator can be very neatly written in terms of shifted kinematics~\cite{Bonocore:2022abn, Balsach:2023ema} that can be chosen to fulfil on-shellness and momentum conservation exactly~\cite{Balsach:2024rkn}.
This is also true for polarised scattering, albeit with some subtleties~\cite{Kollatzsch:2022bqa}.
The soft function $\mathcal{S}$ is similar to the eikonal $\eik$ except that it has a sum over three partons rather than two.

Since all matrix elements are presumed to be known in \mcmule{} it is hence possible to implement~\eqref{eq:nts} by only knowing the momenta and charges of the particles involved without any process-dependent calculations.

\section{Event generation}\label{sec:generator}

The techniques and tools described so far are sufficient to calculate any fixed-order differential distribution.
However, to model experiments with a Monte Carlo, we need to be able to generate events as well.
The naive `garden hose' approach of just storing the sampled momenta and weights is suboptimal.
Since we are using a subtraction scheme, many events will end up with negative weight.
These are not in and of themselves a problem but increase the number of samples required to reach a given statistical error.
A naive estimate would suggest that if $r\times N$ of $N$ weights are negative, we would need $\propto1/(1-2r)^2$ more events to reach the same precision.
If the events are just quickly histogrammed this is not too big of an issue.
However, if, for example, they are propagated through a costly experimental detector simulation, the number of events needs to be minimised.
In practice, this means we need to cancel the negative events early as part of the Monte Carlo rather than late in the histogram.

To do this, \mcmule{} implements a technique called cellular resampling~\cite{Andersen:2021mvw,Andersen:2023cku}.
This method is based on two central ideas
\begin{itemize}
    \item
        integrated cross section are positive, \emph{regardless of the size of the integration domain $\cell$}
        \begin{align}
            \sigma_{\cell} = \int_\cell \D\sigma>0\,,
        \end{align}

    \item
        experiments have finite resolution.
\end{itemize}
By combining these two facts, we arrive at the resampling algorithm of~\cite{Andersen:2021mvw,Andersen:2023cku}
\newcounter{steps}
\begin{enumerate}
    \item
        Generate events.

    \item
        \label{it:seed}
        Pick the event with the most negative weight $w_i<0$ as a seed for a cell $\cell$.

    \item
        Find nearby events and add them to $\cell$.
        This nearest neighbour search can be performed efficiently using vantage-point trees~\cite{Uhlmann:1991vp,Yianilos:1993vp}.

    \item
        Repeat until either:
        \begin{itemize}
            \item
                the cell becomes too large and events in it become resolvable.
                In this case, keep $w_i$ negative

            \item
                the total weight is positive
                \begin{align}
                    \sum_{i\in\cell} w_i > 0\,.
                \end{align}
                In this case, reweight
                \begin{align}
                    w_i \to \frac{\sum_{j\in\cell} w_j}{\sum_{j\in\cell} |w_j|} w_i
                \end{align}
        \end{itemize}

    \item
        Go back to Step~\ref{it:seed} until no more events can be resampled.

    \setcounter{steps}{\value{enumi}}
\end{enumerate}
An animation of this procedure can be found in the supplementary materials to this submission.

If the original number of events was high enough, this algorithm is guaranteed to remove all negative weights without biasing any physical observable.
However, this is often difficult to achieve in practice.
Hence, we extend this algorithm by one more step
\begin{enumerate}
    \setcounter{enumi}{\value{steps}}
    \item
        Generate more events in all cells that still contain a negative weight.

\end{enumerate}
This is possible because we are using cellular resampling as part of the generator rather than as a post-processor for an already existing set of events.

To make use of the cellular resampling algorithm, we need to define a distance in event space.
This metric must ensure that events that differ only by soft photons are considered close to each other (IR safety) and ideally also considers events that are similar to be close to each other.

\begin{figure}[t]
    \centering
    \includegraphics[width=\textwidth]{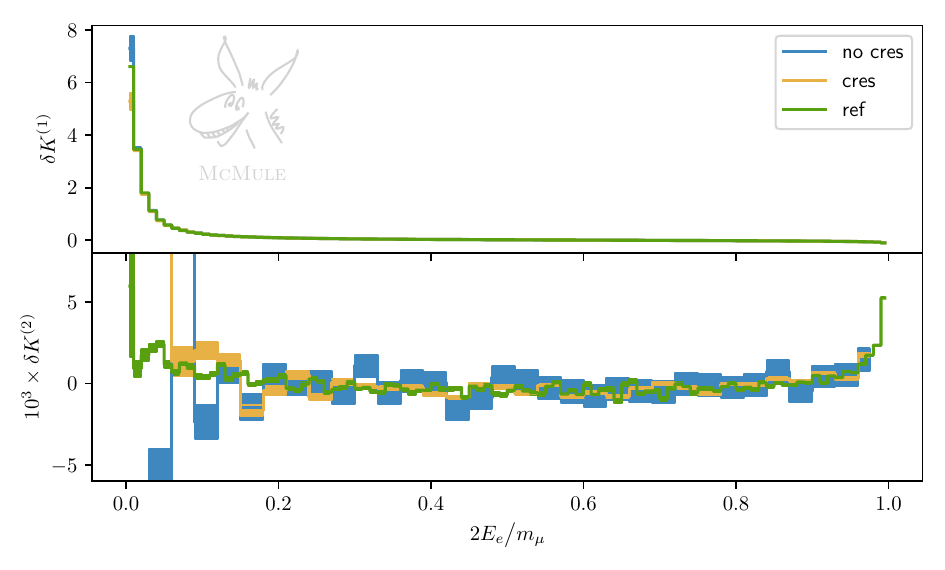}
    \caption{
        The results for the differential $K$ factor $\delta K^{(i)} = (\D\sigma^{(i)}/\D x)\Big/(\D\sigma^{(i-1)}/\D x)$ for the energy fraction of the electron $x=2E_e/m_\mu$ in $\mu\to e\nu\bar\nu$.
        \mcmule{} ran in the default mode (``ref''), event generator mode (``no cres'') and cell-resampling mode (``cres'').
    }
    \label{fig:mudec}
\end{figure}

In the case of $\mu\to\nu\bar\nu e$ where only the electron is detected, a suitable metric might be
\begin{align}
    d(e_1, e_2) = \sqrt{
        \Big|2E_e^{(1)}/m_\mu - 2E_e^{(2)}/m_\mu\Big|^2
        + \Big|\cos\theta_e^{(1)} - \cos\theta_e^{(2)}\Big|^2
    }\,,
\end{align}
though further kinematic information may be added as required.
In Figure~\ref{fig:mudec} we show the size of the corrections for the energy spectrum of the electron at NLO and NNLO both with and without resampling.
The resampling reduces the fraction of negative weights from
\begin{align}
    r \approx 2\times 10^{-2} \to 2\times 10^{-5}\,.
\end{align}
Further, the ``badness'' of the negative weights also gets reduced.
To quantify this we compare the worst weight $w_{\rm min}$ with the average weight $\langle w\rangle$
\begin{align}
    \frac{w_{\rm min}}{\langle w\rangle} \approx -10^5 \to -10^{-3}\,.
\end{align}
This indicates that cellular resampling successfully handles the problem of negative events, allowing \mcmule{} to generate events that are well-suited for experimental studies.

\section{Conclusion}\label{sec:conclusion}

In these proceedings we have discussed the \mcmule{} framework for higher-order QED calculations.
We further introduced the event generation aspect and demonstrated cellular resampling as part of a Monte Carlo tool.
While already well established for NNLO-QED, we plan to extend our code to include more processes such as $e\mu\to e\mu\gamma$ at NNLO, $eN\to eN$ for various nuclei such as carbon or deuterium as well as $ee\to\pi\pi$.
We are also working on the inclusion of electroweak effects for $ee\to ee,\mu\mu$ at NNLO.

\section*{Acknowledgement}
I would like to thank my colleagues in the \mcmule{} Team for their support developing and implementing this framework.

\bibliographystyle{JHEP}
\bibliography{mcmule-acat}

\providecommand{\href}[2]{#2}\begingroup\raggedright\begin{thebibliography}{10}

\bibitem{Banerjee:2020rww}
P.~Banerjee, T.~Engel, A.~Signer and Y.~Ulrich, \emph{{QED at NNLO with
  McMule}}, \href{https://doi.org/10.21468/SciPostPhys.9.2.027}{\emph{SciPost
  Phys.} {\bfseries 9} (2020) 027}
  [\href{https://arxiv.org/abs/2007.01654}{{\ttfamily 2007.01654}}].

\bibitem{Engel:2019nfw}
T.~Engel, A.~Signer and Y.~Ulrich, \emph{{A subtraction scheme for massive
  QED}}, \href{https://doi.org/10.1007/JHEP01(2020)085}{\emph{JHEP} {\bfseries
  01} (2020) 085} [\href{https://arxiv.org/abs/1909.10244}{{\ttfamily
  1909.10244}}].

\bibitem{Banerjee:2021mty}
P.~Banerjee, T.~Engel, N.~Schalch, A.~Signer and Y.~Ulrich, \emph{{Bhabha
  scattering at NNLO with next-to-soft stabilisation}},
  \href{https://doi.org/10.1016/j.physletb.2021.136547}{\emph{Phys. Lett. B}
  {\bfseries 820} (2021) 136547}
  [\href{https://arxiv.org/abs/2106.07469}{{\ttfamily 2106.07469}}].

\bibitem{Broggio:2022htr}
A.~Broggio et~al., \emph{{Muon-electron scattering at NNLO}},
  \href{https://doi.org/10.1007/JHEP01(2023)112}{\emph{JHEP} {\bfseries 01}
  (2023) 112} [\href{https://arxiv.org/abs/2212.06481}{{\ttfamily
  2212.06481}}].

\bibitem{Engel:2023arz}
T.~Engel, F.~Hagelstein, M.~Rocco, V.~Sharkovska, A.~Signer and Y.~Ulrich,
  \emph{{Impact of NNLO QED corrections on lepton-proton scattering at MUSE}},
  \href{https://doi.org/10.1140/epja/s10050-023-01153-x}{\emph{Eur. Phys. J. A}
  {\bfseries 59} (2023) 253}
  [\href{https://arxiv.org/abs/2307.16831}{{\ttfamily 2307.16831}}].

\bibitem{Banerjee:2021qvi}
P.~Banerjee, T.~Engel, N.~Schalch, A.~Signer and Y.~Ulrich, \emph{{M\o{}ller
  scattering at NNLO}},
  \href{https://doi.org/10.1103/PhysRevD.105.L031904}{\emph{Phys. Rev. D}
  {\bfseries 105} (2022) L031904}
  [\href{https://arxiv.org/abs/2107.12311}{{\ttfamily 2107.12311}}].

\bibitem{Kollatzsch:2022bqa}
S.~Kollatzsch and Y.~Ulrich, \emph{{Lepton pair production at NNLO in QED with
  EW effects}},
  \href{https://doi.org/10.21468/SciPostPhys.15.3.104}{\emph{SciPost Phys.}
  {\bfseries 15} (2023) 104}
  [\href{https://arxiv.org/abs/2210.17172}{{\ttfamily 2210.17172}}].

\bibitem{Banerjee:2022nbr}
P.~Banerjee, A.~Coutinho, T.~Engel, A.~Gurgone, A.~Signer and Y.~Ulrich,
  \emph{{High-precision muon decay predictions for ALP searches}},
  \href{https://doi.org/10.21468/SciPostPhys.15.1.021}{\emph{SciPost Phys.}
  {\bfseries 15} (2023) 021}
  [\href{https://arxiv.org/abs/2211.01040}{{\ttfamily 2211.01040}}].

\bibitem{Pruna:2017upz}
G.M.~Pruna, A.~Signer and Y.~Ulrich, \emph{{Fully differential NLO predictions
  for the radiative decay of muons and taus}},
  \href{https://doi.org/10.1016/j.physletb.2017.07.008}{\emph{Phys. Lett. B}
  {\bfseries 772} (2017) 452}
  [\href{https://arxiv.org/abs/1705.03782}{{\ttfamily 1705.03782}}].

\bibitem{Pruna:2016spf}
G.M.~Pruna, A.~Signer and Y.~Ulrich, \emph{{Fully differential NLO predictions
  for the rare muon decay}},
  \href{https://doi.org/10.1016/j.physletb.2016.12.039}{\emph{Phys. Lett. B}
  {\bfseries 765} (2017) 280}
  [\href{https://arxiv.org/abs/1611.03617}{{\ttfamily 1611.03617}}].

\bibitem{Buccioni:2017yxi}
F.~Buccioni, S.~Pozzorini and M.~Zoller, \emph{{On-the-fly reduction of open
  loops}}, \href{https://doi.org/10.1140/epjc/s10052-018-5562-1}{\emph{Eur.
  Phys. J. C} {\bfseries 78} (2018) 70}
  [\href{https://arxiv.org/abs/1710.11452}{{\ttfamily 1710.11452}}].

\bibitem{Buccioni:2019sur}
F.~Buccioni, J.-N.~Lang, J.M.~Lindert, P.~Maierh\"ofer, S.~Pozzorini, H.~Zhang
  et~al., \emph{{OpenLoops 2}},
  \href{https://doi.org/10.1140/epjc/s10052-019-7306-2}{\emph{Eur. Phys. J. C}
  {\bfseries 79} (2019) 866}
  [\href{https://arxiv.org/abs/1907.13071}{{\ttfamily 1907.13071}}].

\bibitem{Yennie:1961ad}
D.R.~Yennie, S.C.~Frautschi and H.~Suura, \emph{{The infrared divergence
  phenomena and high-energy processes}},
  \href{https://doi.org/10.1016/0003-4916(61)90151-8}{\emph{Annals Phys.}
  {\bfseries 13} (1961) 379}.

\bibitem{Frixione:1995ms}
S.~Frixione, Z.~Kunszt and A.~Signer, \emph{{Three jet cross-sections to
  next-to-leading order}},
  \href{https://doi.org/10.1016/0550-3213(96)00110-1}{\emph{Nucl. Phys. B}
  {\bfseries 467} (1996) 399}
  [\href{https://arxiv.org/abs/hep-ph/9512328}{{\ttfamily hep-ph/9512328}}].

\bibitem{Frederix:2009yq}
R.~Frederix, S.~Frixione, F.~Maltoni and T.~Stelzer, \emph{{Automation of
  next-to-leading order computations in QCD: The FKS subtraction}},
  \href{https://doi.org/10.1088/1126-6708/2009/10/003}{\emph{JHEP} {\bfseries
  10} (2009) 003} [\href{https://arxiv.org/abs/0908.4272}{{\ttfamily
  0908.4272}}].

\bibitem{Low:1958sn}
F.E.~Low, \emph{{Bremsstrahlung of very low-energy quanta in elementary
  particle collisions}},
  \href{https://doi.org/10.1103/PhysRev.110.974}{\emph{Phys. Rev.} {\bfseries
  110} (1958) 974}.

\bibitem{Burnett:1967km}
T.H.~Burnett and N.M.~Kroll, \emph{{Extension of the low soft photon theorem}},
  \href{https://doi.org/10.1103/PhysRevLett.20.86}{\emph{Phys. Rev. Lett.}
  {\bfseries 20} (1968) 86}.

\bibitem{Engel:2021ccn}
T.~Engel, A.~Signer and Y.~Ulrich, \emph{{Universal structure of radiative QED
  amplitudes at one loop}},
  \href{https://doi.org/10.1007/JHEP04(2022)097}{\emph{JHEP} {\bfseries 04}
  (2022) 097} [\href{https://arxiv.org/abs/2112.07570}{{\ttfamily
  2112.07570}}].

\bibitem{Engel:2023ifn}
T.~Engel, \emph{{The LBK theorem to all orders}},
  \href{https://doi.org/10.1007/JHEP07(2023)177}{\emph{JHEP} {\bfseries 07}
  (2023) 177} [\href{https://arxiv.org/abs/2304.11689}{{\ttfamily
  2304.11689}}].

\bibitem{Engel:2023rxp}
T.~Engel, \emph{{Multiple soft-photon emission at next-to-leading power to all
  orders}}, \href{https://doi.org/10.1007/JHEP03(2024)004}{\emph{JHEP}
  {\bfseries 03} (2024) 004}
  [\href{https://arxiv.org/abs/2311.17612}{{\ttfamily 2311.17612}}].

\bibitem{Bonocore:2022abn}
D.~Bonocore and A.~Kulesza, \emph{{Next-to-leading power corrections for soft
  photon bremsstrahlung}},
  \href{https://doi.org/10.22323/1.414.1128}{\emph{PoS} {\bfseries ICHEP2022}
  (2022) 1128}.

\bibitem{Balsach:2023ema}
R.~Balsach, D.~Bonocore and A.~Kulesza, \emph{{Soft-photon spectra and the LBK
  theorem}},  \href{https://arxiv.org/abs/2312.11386}{{\ttfamily 2312.11386}}.

\bibitem{Balsach:2024rkn}
R.~Balsach, A.~Kulesza and D.~Bonocore, \emph{{The Emission of Soft-photons and
  the LBK Theorem, Revisited}},
  \href{https://doi.org/10.5506/APhysPolBSupp.17.2-A8}{\emph{Acta Phys. Polon.
  Supp.} {\bfseries 17} (2024) 2}
  [\href{https://arxiv.org/abs/2401.01820}{{\ttfamily 2401.01820}}].

\bibitem{Andersen:2021mvw}
J.R.~Andersen and A.~Maier, \emph{{Unbiased elimination of negative weights in
  Monte Carlo samples}},
  \href{https://doi.org/10.1140/epjc/s10052-022-10372-3}{\emph{Eur. Phys. J. C}
  {\bfseries 82} (2022) 433}
  [\href{https://arxiv.org/abs/2109.07851}{{\ttfamily 2109.07851}}].

\bibitem{Andersen:2023cku}
J.R.~Andersen, A.~Maier and D.~Ma\^\i{}tre, \emph{{Efficient negative-weight
  elimination in large high-multiplicity Monte Carlo event samples}},
  \href{https://doi.org/10.1140/epjc/s10052-023-11905-0}{\emph{Eur. Phys. J. C}
  {\bfseries 83} (2023) 835}
  [\href{https://arxiv.org/abs/2303.15246}{{\ttfamily 2303.15246}}].

\bibitem{Uhlmann:1991vp}
J.K.~Uhlmann, \emph{Satisfying general proximity/similarity queries with metric
  trees}, \href{https://doi.org/10.1016/0020-0190(91)90074-R}{\emph{Inf.
  Process. Lett.} {\bfseries 40} (1991) 175}.

\bibitem{Yianilos:1993vp}
P.N.~Yianilos, \emph{Data structures and algorithms for nearest neighbor search
  in general metric spaces},  in \emph{Proceedings of the Fourth Annual
  ACM-SIAM Symposium on Discrete Algorithms}, SODA '93, (USA), p.~311–321,
  Society for Industrial and Applied Mathematics, 1993.

\end{thebibliography}\endgroup

\end{document}